\documentstyle[12pt]{article}
\begin{document}
\title{Determination of two-body potentials from $n$-body spectra}
\author{
E.J.O. Gavin \\
 NIKHEF-K,  P.O. Box 41822, NL-1009 DB Amsterdam,
\\ The Netherlands
\\ \\
H. Leeb \\
  Institut f\"{u}r Kernphysik, Technische Universit\"{a}t Wien,
  \\ Wiedner Hauptstr. 8-10, A-1040 Wien,
Austria \\
\\ H. Fiedeldey \\ Physics Department, University of South
Africa,\\
 P.O. Box 392,
0001 Pretoria, South Africa
}
\vspace{0.65in}
\date{}
\maketitle
\vspace{1.0in}
\begin{abstract}
 We show how the two-body potential may be uniquely determined from
$n$-body spectra in cases where the hypercentral approximation
is valid.  We
illustrate this by considering an harmonic oscillator potential
which has
been altered by changing the energy or normalisation constant of
the ground
state of the $n$-body system and finding how this modifies the
two-body
potential. It is  shown that with increasing number of particles
the
spectrum must be known more precisely to obtain the  two-body
potential
to the same degree of accuracy.
\end{abstract}

\newpage

\section{Introduction}
The standard inverse scattering theory is formulated for two-body
systems \cite{inv}. Its aim is the determination of the interaction
from the spectral information pertaining to bound and scattering
states of the two particles. In some cases there is no
or insufficient spectral
information available on the two-body system, while there is
access to the spectral data of a corresponding many-body
system. An example of this is the baryon spectrum whereas
no two-quark system is known to exist.
Thus   it is equally interesting to ask whether it is possible
to extract the two-body potential from the bound and scattering
states of a many-body system. To our knowledge this general
question has never been considered. There is a schematic treatment
of the three-body problem in one space
dimension by Zakhariev \cite{thr}, but this is  unsuitable for
practical applications. With respect
to applications, only recently \cite{Leeb} has a feasible method for
the inversion of baryon spectra  been proposed and have calculations
of the associated quark-quark potentials been reported.

In the following, spectral information given in terms of
many-body bound states is  considered. For many-body bound
states there are approximations which bear a formal resemblance
to a two-body Schr\"{o}dinger equation. This property suggests
the possibility of applying standard two-body inversion
techniques within these approaches.
Examples of these include the Hartree-Fock method in which the
interaction
of the bodies is described by a mean field, and
the hypercentral approximation (HCA) in which only the hyperradial
 part of the interaction in the
 hyperspherical harmonic expansion method
\cite{HHEM} (HHEM) is retained. It is on the latter scheme,
described in more detail in the following section,
that we will focus in this article. The hypercentral potential
used in this method is determined uniquely by the underlying
two-body interaction. The main progress achieved in this study
is an inversion of this relation for a
general $n$-body system consisting of $n$ identical
 particles. This inversion procedure
is formulated for arbitrary $n$ and can be
applied as long as the HCA gives a proper description of
the many-body system.
In principle the method is an extension of the
inversion procedure of baryon spectra, investigated in
previous papers \cite{Leeb,Gavin}. In this extended scheme we
 show how the two-body interaction  may be obtained in
 general from  $n$-body spectral
information, and in particular use the method to obtain the
two-body
interaction from  three-, four-, five- and six-body spectral data.

We illustrate this scheme by applying our inversion method
to spectra which coincide with that of the harmonic oscillator
in both energy levels, and normalisation constants of
the bound states in hyperspherical space, bar the ground state.
This ground state is either shifted in energy relative to the
harmonic oscillator ground state or has a different bound state
normalisation constant than that of the harmonic oscillator ground
state.  Primarily we focus on the changes of the two-body
potential with increasing number of particles $n$.
Our example clearly demonstrates that the accuracy of the
determination of the two-body potential from the $n$-body spectrum
deteriorates with increasing $n$, if the  accuracy to which the
bound states are given remains fixed. It implies a rapid loss of
information on the underlying two-particle potential with
increasing size of the bound system of particles, despite the
corresponding increase in the number of states present in their
spectrum. This can also be turned around and viewed as an
indication that the $n$-body spectrum becomes less sensitive to
the details of the two-body interaction with increasing number
of particles.

The details of the inversion in the HCA are described in the
following
section, and the illustration and the results are to be found in
section 3. We conclude in section 4, discussing what may be learnt
from the example we have used.

\section{The hypercentral approximation and inversion formalism}
We begin this section with a description of the HCA and in
particular show the relation of the hypercentral potential to
Wigner-type two-body interactions. We then show how this relation
may be inverted, thereby facilitating an inversion of  $n$-body
spectral data to obtain the underlying two-body interaction.

\subsection{The hypercentral approximation}
An $n$-body state is described after removing the centre-of-mass
motion by 3($n$-1) internal co-ordinates.
The corresponding $n$-particle
Schr\"{o}dinger equation can generally only be solved using various
approximations.
 A very effective
procedure is the expansion of the wave functions and potentials
in terms of hyperspherical harmonics. This HHEM is particularly
useful for the evaluation of many-body bound states of both bosons
and fermions.

 Specifically
for fairly soft two-body interactions, where correlations between
the particles are negligible, the lowest order of the HHEM, the
so-called hypercentral approximation (HCA), provides an excellent
description of the many-body bound states.
It is exact in the case of the $n$-body harmonic oscillator, where
it is well know that the $n$-body bound states may be obtained as
linear combinations of single particle oscillator eigenfunctions.
A prominent example of the applicability of the HCA is the
calculation of baryon spectra \cite{Richard} in the framework of the
non-relativistic quark model, in which the forces operating
between the constituent quarks are soft and confining.
Even for a Coulomb interaction or certain nucleon-nucleon
interactions, the HCA alone may account for up to 90 \% of the
binding energy of an $n$-body bound state of the system (see
\cite{HHEM} and references therein).

In the HCA the $n$-body Schr\"{o}dinger equation reduces to an
effective Schr\"{o}dinger equation in the hyperradius $\rho$
\begin{equation}
 \rho ^{2} = \frac{2}{n} \sum_{i < j} ^{n} ({\bf r}_i - {\bf r}_j)^2,
\end{equation}
where ${\bf r}_i$ denotes the co-ordinate of the $i$-th particle.
The hypercentral potential $V^n_{hc}(\rho)$ which accounts for the
 interactions of the  $n$ bodies
is the
rotationally invariant part of the sum of two-body central potentials
$U ( \left| {\bf r}_i - {\bf r}_j \right| )$
acting between each pair of particles $i$ and $j$ in the
$3(n-1)$-dimenional space. It is determined by the two-body
potential
$U$ through the linear relation \cite{HHEM,Fermion}
\begin{eqnarray}
V^n_{hc} (\rho) &  =  & \sum_{j=0}^{2l_m} V^n_j (\rho) \nonumber \\
  &  = & \sum_{j=0}^{2 l_m} \frac{a_j}{h}
 \int^{+1}_{-1} dz ~ (1-z)^{\frac{D-5}{2}
+ L_m - j} (1+z)^{j + \frac{1}{2}} ~ U(\rho \sqrt{\frac{1+z}{2}} ) ,
\label{jvhc}
\end{eqnarray}
where $D$ has been used to denote $3 (n-1)$, and $l_m$ is the maximum
value of the orbital angular momentum in the outermost shell
consistent
with a hyperspherical harmonic of degree $L_m$. By allowing
the minimal degree $L_m$ of hyperspherical harmonics present in the
expansion of the $n$-body wavefunction to be greater than zero, the
Pauli Principle may be implemented for bound states of fermions.
The normalisation
constant $h$ is
 defined by
\begin{equation}
h = \sum_{j=0}^{2 l_m} a_j \int^{+1}_{-1} dz (1-z)^{\frac{D-5}{2}
+ L_m -j} (1+z)^{j+\frac{1}{2}} .
\end{equation}
The $a_j$ are constants determined by the shell structure of the
bound state.
If $L_m = 0$, as is the case for all $n$-boson bound states,
the relation simplifies to
\begin{equation}
V^n_{hc}(\rho)
 = \frac{\Gamma (\frac{D}{2} )}{\sqrt{\pi} 2^{D/2 - 2}
   \Gamma ( \frac{D-3}{2} )} \int _{-1} ^{+1} dz (1 - z)^
{\frac{D-5}{2}} (1+z)^{\frac{1}{2}} U (\rho \sqrt{\frac{1+z}{2}} ) .
\label{vhc}
\end{equation}
Thus $V^n_{hc}(\rho )$
represents
some weighted average of the two-body potential over the range
$(0,\rho)$,
reflecting the range of the possible separations of pairs of particles
in the $n$-body state of hyperradius $\rho$.  The
total $n$-body interaction due to the$\frac{n(n-1)}{2}$
 pairs of particles interacting via the two-body interaction $U$ is
then accounted for by  $\frac{n(n-1)}{2}
V ^n _{hc} (\rho)$.
For $n$ particles of equal mass $m$, a bound state with energy $E$ is
determined in the HCA by the solution of
\begin{equation}
\left\{ \frac{\hbar ^2}{m} \left[ - \frac{d^2}{d \rho ^2} +
 \frac { {\cal L} ( {\cal L} + 1) }{\rho ^2} \right]
+ \frac{n(n-1)}{2} V ^n _{hc} (\rho) - E \right\} u(\rho ) = 0
\label{hca}
\end{equation}
where the so-called grand orbital momentum ${\cal L}$ depends on the
orbital angular momentum $l$ and the dimension of the space,
\begin{equation}
{\cal L} = l + L_m + \frac{D-3}{2} .
\end{equation}
For all two-body potentials which are less singular than $r^{-3}$
at the origin, the hypercentral potential is well-defined. This implies
that the hypercentral potential is also
 no more singular than this, as the leading order behaviour
of the hypercentral potential is that of the two-body potential
near the orgin. This condition does
not in reality restrict the use of the HCA, as an attractive potential
must be less singular than $r^{-2}$ if the spectrum is to be bounded
from below, while the presence of a repulsive singularity in the
potential would signal the possible inappropriateness of the HCA,
as correlations between the
particles could be important.

In physical systems where the HCA provides a good description
of the spectrum, $n$-body spectral data could be used via
standard inversion techniques to obtain the effective interaction
\begin{equation}
 \frac{{\cal L} ({\cal L} + 1)}
{\rho ^2} + \frac{m}{\hbar ^2} \frac{n(n-1)}{2} V^n_{hc} (\rho)
\label{eff}
\end{equation}
appearing in equation (\ref{hca}).  The corresponding two-body
potential may then be deduced from the hypercentral potential via
a further inversion step.

In the following we present such a procedure.
As the expression is more
complex if $L_m > 0$, we first derive the relation for an $n$-boson
hypercentral potential, and then use these results to generalise to
the $n$-fermion case.

\subsection{The inversion of the $n$-boson hypercentral potential}
Here we consider the inversion of the hypercentral potential in
the simplest case, where $L_m = 0$.
Using the new
variables
\begin{equation}
x = \rho ^2 \quad\mbox{and}\quad y =
\rho ^2 \left( \frac{1 + z}{2} \right) ,
\label{xy}
\end{equation}
we  rewrite the relation (\ref{vhc}) in a form which we can
readily invert to obtain the two-body potential $U$ in
terms of the hypercentral potential $V^n_{hc}$,
\begin{equation}
\frac{ \sqrt{\pi} \Gamma \left( \frac{D-3}{2} \right) }{2 \Gamma
\left( \frac{D}{2} \right) } x^{\frac{D-2}{2}} V^n_{hc} ( \sqrt{x} )
= \int^{x}_{0} dy \sqrt{y} (x-y)^{\frac{D-5}{2}} U (\sqrt{y} ).
\label{difform}
\end{equation}
We consider two distinct cases, depending on whether the system
consists of an odd or even number of particles.
\begin{enumerate}
\item If $n$ is even, we may simply differentiate the above
expression to
obtain the two-body potential in terms of the hypercentral potential.
Setting $D-5 = 2k$, where $k=2,5,8,...$, we find
\begin{equation}
U(\sqrt{y}) = \frac{\sqrt{\pi}}{2} \frac{1}{\Gamma
\left( k + \frac{5}{2}
 \right)} \frac{1}{
\sqrt{y} } \frac{d^{k+1}}{dy^{k+1}} \left[ y^{k + \frac{3}{2}}
V^n_{hc} (\sqrt{y}) \right].
\label{neven}
\end{equation}

\item If $n$ is odd, we write $D-5 = 2k+1$, where $k=0,3,6,...$,
and on
differentiating expression (\ref{difform}) we obtain
\begin{equation}
\frac{d^{k+1} }{dx^{k+1}} \left[ x^{k+2} V^n_{hc} (\sqrt{x}) \right]
= \frac{2~\Gamma \left(
k+3 \right) }{ \pi
 } \int^{x}_{0} dy
\frac{ \sqrt{y} ~ U (\sqrt{y})}{\sqrt{x-y}}.
\label{nodd}
\end{equation}
This is in the form of the Abel integral equation \cite{Abel},
\begin{equation}
g(s) = \int^{s}_{0} dt \frac{f(t)}{\sqrt{s-t}}
\end{equation}
which may be uniquely inverted to obtain
\begin{equation}
f(t) = \frac{1}{\pi} \int^{t}_{0} ds \frac{dg}{ds}
\frac{1}{\sqrt{t-s}}
+ \frac{1}{\pi} \frac{g(0)}{\sqrt{t}}.
\end{equation}
The Abel transform is a special case of the Riemann-Liouville
fractional integral \cite{Erdelyi}.
Thus we can invert (\ref{nodd}) to obtain
\begin{eqnarray}
U(\sqrt{y}) &  = &  \frac{1}{2}
\frac{1}{\Gamma \left( k + 3 \right)} \frac{1}{ \sqrt{y} }
 \\   &  &  \left[ \int^{y}_{0}
\frac{d^{k+2}}{dx^{k+2}} \left[ x^{k+2} V^n_{hc} (\sqrt{x}) \right]
\frac{dx}{\sqrt{y-x}}
    + \frac{1}{\sqrt{y}} \lim_{x \rightarrow 0}
\frac{d^{k+1}}{dx^{k+1}} \left[ x^{k+2} V^n_{hc} (\sqrt{x}) \right]
\right] . \nonumber
\end{eqnarray}
 The
derivation of the two-body potential from the three-body
hypercentral
potential presented previously \cite{Leeb,Gavin} is an example
of the
above transformation with $k=0$.

\end{enumerate}

While these relations above show that it is in principle possible to
extract the two-body interaction from a hypercentral potential
describing a system of arbitrary $n$,
it is clear from the form of these relations that in practice
the larger the number of particles, the more difficult it is to
perform this inversion numerically as higher order derivatives enter
the relation.  We will demonstrate this  quantitatively in the
following section, following the generalisation to the $n$-fermion
case.

\subsection{The inversion of the $n$-fermion hypercentral potential}

Using the variables $x$ and $y$ as defined in(\ref{xy}), we may
 re-express each
$V^n_j(\rho)$ appearing in (\ref{jvhc})
as
\begin{equation}
V^n_j (\sqrt{x}) = \frac{a_j}{h} \left( \frac{2}{x} \right)
 ^{\frac{D-5}{2}
+ L_m + \frac{3}{2}} \int^x_0 dy (x-y)^{\frac{D-5}{2} + L_m - j }
y^{j+\frac{1}{2}}
U (\sqrt{y}) .
\end{equation}
Then $V^n_j$ may be found in terms of the two-body potential as
described in the preceeding section.

If there are an even number of fermions, setting $D-5 = 2k$, where
 $k =$ 2,5,8,..., we find
\begin{equation}
A_j y^{j+\frac{1}{2}} U(\sqrt{y}) = \frac{d^{k+L_m-j+1}}{dy
^{k+L_m-j+1}} \left[ y^{k+L_m + \frac{3}{2}} V^n_j (\sqrt{y})
\right],
 ~ j = 0,1,...2 l_m,
\end{equation}
where we have defined the constant $A_j$ as
\begin{equation}
A_j = 2^{k+L_m + \frac{3}{2}} (k+L_m - j)! ~ \frac{a_j}{h}.
\end{equation}
Taking the $j^{th}$ derivative with respect to $y$ of each of the
above
expressions and then summing over $j$ we find the relationship
from which
$U$ may be determined from a known hypercentral potential:
\begin{equation}
\sum^{2 l_m}_{j=0} A_j \frac{d^j}{dy^j} \left[ y^{j + \frac{1}{2}}
U(\sqrt{y}) \right] = \frac{d^{k+L_m+1}}{dy^{k+L_m + 1}}
\left[ y^{k+L_m+\frac{3}{2}} V^n_{hc} (\sqrt{y} ) \right] .
\end{equation}

For an odd number of fermions, writing $D-5 = 2k+1$,
where $k =$ 0,3,6,...,
we have
\begin{equation}
\frac{d^{k+L_m-j+1}}{dx^{k+L_m-j+1}} \left[ x^{k+L_m+2}
V^n_j(\sqrt{x})
\right] = B_j \int^x_0 dy \frac{y^{j + \frac{1}{2}} U(\sqrt{y}) }
{\sqrt{y-x}}
\end{equation}
where the constant $B_j$ is defined as
\begin{equation}
B_j = \frac{\Gamma \left( k + \frac{3}{2} + L_m - j \right)
2^{k + \frac{3}{2} + L_m}}{\sqrt{\pi}}   ~ \frac{a_j}{h}.
\end{equation}
If $V^n_{hc}(x)$ is no more singular than $x^{-2}$, then use of the
Abel transform once again leads to the expressions
\begin{eqnarray}
B_{0} y^{\frac{1}{2}} U(\sqrt{y}) &  = & \frac{1}{\pi}
\int^{y}_{0} \frac{d^{k+L_m+2}}
{dy^{k+L_m+2}} \left[x^{k+L_m+2} V^n_0 (\sqrt{x}) \right] \frac{dx}
{\sqrt{y-x}} \nonumber \\  &  &
 + \frac{1}{\pi \sqrt{y}} \lim_{x \rightarrow 0} \frac{d^{k+L_m+1}}
{dx^{k+L_m+1}} \left[ x^{k+L_m+2} V^n_0 (\sqrt{x}) \right]
\end{eqnarray}
and
\begin{equation}
B_j y^{j + \frac{1}{2}} U(\sqrt{y}) = \frac{1}{\pi} \int^y_0
\frac{d^{k+L_m-j+2}}{d^{k+L_m-j+2}} \left[ x^{k+L_m+2} V^n_j
(\sqrt{x})
\right] \frac{dx}{\sqrt{y-x}}, ~~ j = 1,2,...2 l_m.
\end{equation}
Taking derivatives with respect to $y$ and performing an integration
by parts we find
\begin{eqnarray}
B_j \frac{d^j}{dy^j} \left[ y^{j + \frac{1}{2}} U(\sqrt{y}) \right]
 & = & \frac{1}{\pi} \int^y_0 \frac{d^{k+L_m+2}}{dx^{k+L_m+2}}
\left[x^{k+L_m+2} V^n_j (\sqrt{x}) \right] \frac{dx}{\sqrt{y-x}}
\nonumber \\
 &   +  & \frac{1}{\pi} \frac{1}{ \sqrt{y-x}}
\frac{d^{k+L_m+1}}{dx^{k+L_m+1}}
\left[ x^{k+L_m+2} V^n_j(\sqrt{x}) \right]|_{x=0},
j = 1,2,...2 l_m .
\end{eqnarray}
Summing these expressions over $j$ we obtain a differential equation
\begin{eqnarray}
\sum_{j=0}^{2 l_m} B_j \frac{d^j}{dy^j} \left[y^{j+\frac{1}{2}} U(
\sqrt{y}) \right] & = & \frac{1}{\pi}
\int^y_0 \frac{d^{k+L_m+2}}{dx^{k+L_m
+2}} \left[ x^{k+L_m+2} V^n_{hc} (\sqrt{x}) \right]
\frac{dx}{\sqrt{y}}
\nonumber \\  &  + &
\frac{1}{\pi \sqrt{y}} \frac{d^{k+L_m+1}}{dx^{k+L_m+1}} \left[
x^{k+L_m+2} V^n_{hc} (\sqrt{x}) \right] |_{x=0}
\end{eqnarray}
as for the even number of fermions.

Thus in order to obtain the underlying two-body potential from an
$n$-fermion hypercentral potential,
a differential equation of order $2 l_m$
must be solved. Note that if $L_m = 0$, as we have not considered
non-Wigner type
interactions,
 the $n$-fermion result coincides
with that of the $n$-boson result. We may have $L_m = 0$ if we are
dealing
with fermions with some additional degree of freedom such as
isospin or
colour.  Examples of such systems include bound states of three quarks
or three or four nucleons.

\section{The example of an-almost-harmonic oscillator}

The two step procedure, introduced in the previous section, allows
the determination of the two-body potential from the knowledge of
the spectral information of the $n$-body system. In principle this
spectral information for bound states includes both the energy
levels and the normalisation constants, which represent
additional information on the wave functions. Both quantities are
required to fully determine the potential. Usually the energy
levels, which are directly related to the mass spectrum of the bound
system, can be measured with high accuracy. The situation is
less satisfactory for the normalisation constants, which are not
directly accessible to experiment and can only be deduced from
transition observables. Consequently there is a considerable
uncertainty in the normalisation constants.

It is precisely this situation which we want to study in a first
schematic example. Given this new $n$-body inversion procedure, we
would like to have an idea of what may be learnt about  two-body
interactions from
  $n$-body spectral
data. To this end, we perform an inversion of $n$-body spectral
information to obtain the two-body interaction
in the simplest case, where  $L_m = 0$.
As a basic model then, we consider a bound system
of $n$ bosons  of equal mass $m$ interacting via two-body
harmonic oscillator potentials
$\frac{r^2}{b^4}$.
This system
has only a discrete spectrum and is exactly described by the HCA.
The corresponding hypercentral potential is also a harmonic
oscillator potential, $V^n_{hc}=(n-1)^{-1} U$, and the n-body
bound state is obtained by solving a Schr\"{o}dinger-type equation
in the hyperradius $\rho$
\begin{equation}
\left\{ - \frac{d^{2}~}{d \rho ^{2}} + \frac {{\cal L}
( {\cal L} + 1)} {\rho^{2}}
  + \frac{\rho^{2}}{b_n^{4}} - \epsilon \right\} u (\rho) = 0 .
\end{equation}
The energy of this state is $\epsilon \frac{m}{\hbar ^2}$ and the
 potential term is simply given by
\begin{equation}
V = \frac{m}{\hbar ^2} \frac {n(n-1)}{2} V^n_{hc} =
\frac{\rho^{2}}{b_n^{4}}.
\end{equation}
The solution  is of the form
\begin{equation}
u ( \rho ) = \left( \frac{\rho}{b_n} \right) ^{{\cal L} + 1} \exp
\left(
\frac{- \rho ^{2}}{2 b_n^{2}} \right) ~
 _{1}F_{1}(\alpha , \beta , \left( \frac{\rho}{b_n} \right) ^{2} )
\label{psi} ,
\end{equation}
where $_{1}F_{1}$ is the confluent hypergeometric  function of
the arguments
\begin{equation}
\alpha = (2{\cal L} + 3 - \epsilon b_n^{2} )/4
\end{equation}
and
\begin{equation}
\beta = {\cal L} + \frac{3}{2}.
\end{equation}
If $\alpha = -n$, where $n$ is an integer,
one obtains a bound state and (\ref{psi})
becomes
\begin{equation}
u (\rho ) = N \left( \frac{\rho}{b_n} \right) ^{{\cal L} + 1} \exp
\left(
\frac{-{\rho}^{2}}{2 b_n^{2}} \right)
L ^{\beta - 1} _{n} \left( \frac{\rho}{b_n} \right) ^{2} ,
\end{equation}
where $N$ is a constant normalising the wave function,
and $L ^{\beta - 1}_{n}$ denotes a Laguerre polynomial. In
particular, if $\alpha = 0$, corresponding to $\epsilon =
\epsilon
_{0} = (2 {\cal L} + 3)/b_n^{2}$, we obtain the ground state $u_{0}$
\begin{equation}
u_{0} = \left[ \frac{b_n}{2} \Gamma \left( {\cal L} + \frac{3}{2}
\right) \right] ^{- \frac{1}{2}}
\left( \frac{ \rho }{b_n} \right) ^{{\cal L} + 1} \exp \left(
\frac{- \rho ^{2}}{2 b_n^{2}} \right) .
\end{equation}

It is not our intention to reconstruct the two-body harmonic
oscillator potential from the corresponding $n$-body bound-state
spectrum because this can be done analytically. Rather, we want
to simulate the realistic situation taking into account the
uncertainties in the normalisation constants or energies.
Therefore we
study the variations of the two-body potential due to slight
changes in the $n$-body spectrum. In order to demonstrate the
essential points in a simple way we restrict ourselves to small
modifications of the $n$-body ground states. We alter either the
ground state normalisation constant leaving the energy unchanged,
or the ground state energy, keeping the same ground state
normalisation constant.
All
other remaining spectral data of the $n$-body system are
unchanged from those of the $n$-body harmonic oscillator.

Because of our well defined reference potential it is not
necessary to invert this modified $n$-body spectrum via a
standard inversion procedure in order to obtain the corresponding
hypercentral potential. Techniques of supersymmetric quantum
mechanics \cite{Sukumar} offer a more elegant way of achieving
the same result in our case. This method is directly related
to the standard inversion procedure of Gel'fand-
Levitan \cite{Abraham} and has the additional advantage
that we can really take into account the whole $n$-body
spectrum. The latter
point will be very important for the interpretation of our final
result because we can be sure that all variations found in the
two-body potential are generated by the modification of the
$n$-body ground-state normalisation constant. There will be no
artifacts due to the neglect of $n$-body states, which is
unavoidable in a direct application of standard inversion
methods.

We alter the $n$-body spectrum via supersymmetric transformations
in two steps, first, by removing the harmonic
oscillator ground state,
and then by inserting a new ground state into the $n$-body spectrum.
To remove the ground state from the $n$-body spectrum, we perform
the transformation
\begin{equation}
\tilde{V} = V - 2 \frac{d^{2}~}{d\rho^{2}} \ln u_{0}.
\label{remove}
\end{equation}
This results in  the new potential
\begin{equation}
\tilde{V} = \frac{\rho^{2}}{b_n^{4}} + \frac{2}{b_n^{2}}
+ \frac{2({\cal L} + 1)} {\rho^{2}} .
\end{equation}
The solutions $\tilde{u}$ associated with this new potential may
be obtained from equation (\ref{psi}) with the replacements
${\cal L} \rightarrow {\cal L} + 1$ and $\epsilon \rightarrow
\epsilon - \frac{2}{b_n^{2}}$, thus
\begin{equation}
\tilde{u} (\rho) = \left( \frac{\rho}{b_n} \right) ^{{\cal L} + 2}
\exp \left( - \frac{\rho^2}{2 b_n^2} \right) ~_{1} F _{1} (\alpha + 1,
\beta + 1,
\left( \frac{\rho}{b_n} \right) ^{2} ).
\end{equation}
Adding a new state to the spectrum is accomplished by the
transformation
\begin{equation}
\tilde {\tilde{V}} = \tilde{V} - 2 \frac{d^{2}~}{d\rho^{2}} \ln
\tilde{u}
( 1 + \lambda \int^{\infty}_{\rho} \tilde{u}^{-2}(z) dz )
\label{lambda}
\end{equation}
where $\lambda$ is a positive constant which controls the bound state
normalisation constant of the new $n$-body ground state.

We need to check that  the behaviour of $\tilde{\tilde{V}}$
near the origin does not violate the conditions necessary for
construction of the corresponding two-body potential.
Expanding $\tilde{\tilde{V}}-\tilde{V}$ to
leading- and next-to-leading order for small $\rho$, we find
that the change in the potential is
\begin{equation}
-2 ( \ln \tilde{u} ) ''- 2
 (\ln (1 + \lambda \int^{\infty}_{\rho}\tilde{u} ^{-2}(z) dz ))''
= - \frac{2 {\cal L} + 2} {{\rho}^2} -  \frac{4}{b_n^2}
( \frac{\alpha + 1}{\beta + 1} - \frac{1}{2} ) (1 - 2
 \frac {2{\cal L} + 3}{ 2{\cal L} + 1}) + O(\rho^3) ,
\end{equation}
where $'$ denotes differentiation with respect to $\rho$
The $\rho^{-2}$ term can immediately be seen to cancel the
singularity which arises from the removal of the original
ground state.

If we insert a state at the same energy as the ground
state of the harmonic oscillator, this corresponds to
choosing $\tilde{u}$ with $\alpha = 0$, and the constant in
the above expression becomes
\begin{equation}
-  \frac{4}{b_n^2}
( \frac{1}{\beta + 1} - \frac{1}{2} ) (1 - 2
 \frac {2{\cal L} + 3}{ 2{\cal L} + 1}) = -\frac{4}{b_n^2}
\frac{-{\cal L}
- \frac{1}{2}} {2{\cal L} + 5} \frac{-2{\cal L} - 5}{2{\cal L} + 1}
= - \frac{2}{b_n^2}
\end{equation}
which leaves the potential unchanged at the origin. A change
in the energy of the ground state will however produce a
shift in the potential at the origin.

If one merely wishes to change the normalisation constant
of the ground state one can choose $\tilde{u} = u_{0}^{-1}$
\cite{Sukumar} and the
 transformations (\ref{remove}) and (\ref{lambda}) combine to
\begin{equation}
\tilde{\tilde{V}} = V - 2 (ln (1 + \kappa \int^{\rho}
_{0}u_{0}^{2}(z) dz))'' ,
\end{equation}
where use has been made of the fact that $u_{0}$ is the normalized
ground state, and the constant $\kappa$ is related to the $\lambda$ of
the transformation shown in equation (\ref{lambda}) by
\begin{equation}
\kappa = \frac{ - \lambda}{\lambda + 1}.
\end{equation}
Here the relation between the wave function of the new ground state
and that of the original ground state near the origin is determined
by the choice of $\kappa$ \cite{Sukumar}
\begin{equation}
\lim _{\rho \rightarrow 0} \tilde{u} (\rho) = \sqrt{1 + \kappa}
\lim _{\rho \rightarrow 0} u_0 (\rho).
\end{equation}

In our first numerical study we focus on the sensitivity of the
hypercentral and corresponding two-body potential to the $n$-body
spectral data at a fixed particle number $n$ and orbital angular
momentum quantum number $l=0$. Using the four-body case as an
example, we vary either the size of the normalised four-body
ground state wavefunction at the origin, or  shift the energy by
$\pm 5 \%$ and $\pm 10 \%$. The changes induced in the potential
are depicted in Figure 1. While the change in the potential gets
larger as the change in the energy or normalisation constant
increases, it is
obvious that even for these relatively small changes
 the potential does not respond linearly.

We turn our attention to the relative changes in potential that
occur if
we make the same change to the spectra of different number of bodies.
In Figures 2(a) and (b) we show the change in $V^n_{hc}$ induced by a
reduction of the bound state normalisation constant by 5 \% and the
lowering of the ground state energy by 5 \% without change in the
normalisation constant respectively for three, four, five and
six bodies. The corresponding changes induced in the two-body
potentials appear in Figures 3(a) and (b).
For comparison, we include the results in the case of a bound state of
two bodies.
In Figures 3(c) and (d) we show the full new two-body potentials, to
illustrate that the change to the oscillator potential is not so
drastic as to invalidate the use of the HCA.

Turning first to the changes in $V^n_{hc}$ we see that the greater the
number of bodies, the further from the origin the maximum change in
the potential occurs.  This is to be expected from
the form of the HCA:
with an increasing number of bodies, the effective centrifugal barrier
as determined by ${\cal L}$ increases, and sensitivity to the
potential
in this region is reduced. The change in $V^n_{hc}$ decreases as $n$
increases, which provides hope for the possibility of obtaining a
reliable hypercentral potential from the inversion of $n$-body
spectral
data. However, this decrease is not transmitted to the underlying
two-body interaction, as becomes clear on scrutiny of Figure 2.
Here the change induced in the two-body potential is seen to
increase rapidly with increasing particle number, even though
the change
in the corresponding hypercentral potential is decreasing.

Also worth noting, is that the number of oscillations in the
two-body
potential increases with increasing number of bodies. From the
form of
the transformation (\ref{neven}), it is clear that $p$ maxima and
minima in the hypercentral potential translate into $p+k+1$
extrema
in the two-body potential. This has obvious implications for a
potential
obtained through inversion, which may have spurious numerically
induced
oscillations due to the neglect of part of the original $n$-body
spectrum.
The results clearly indicate the limitations on any $n$-body
inversion.

\section{Conclusions}
We have derived a relation for the exact inversion of the underlying
two-body potential from a given $n$-body hypercentral potential.
In principle this implies that in physical systems which are
adequately described by the hypercentral approximation, we have
obtained an exact solution of the $n$-body inverse spectral
problem in terms of two-body forces.

We have studied the possibility of extracting information on the
two-body potential from $n$-body spectral data in the schematic
example of an-almost-harmonic oscillator. It turns out that with
increasing number of particles the $n$-body state becomes less
sensitive to details of the underlying two-body interaction. In
other words a small uncertainty in the $n$-body spectral data
translates into a dramatic change in the two-body potential.
This is especially true of any uncertainty in the bound state
normalisation constant.

The harmonic oscillator is a fairly good approximation to some
realistic systems, e.g. basic features of light nuclei can be
obtained within oscillator models. We are therefore  justified in
 transposing our results to realistic systems. In particular
they imply that the accuracy of the bound state measurements
for larger systems must  increase considerably  if we want to
get comparable information on the two-body potential. This
situation also corresponds to an intuitive understanding of the
increasing difficulty of extracting the contribution to the
interaction of a pair of particles, from the total interaction
of the $\frac{n(n-1)}{2}$ pairs, which has been
accounted for only in an average way. From this point of view it
appears to be a further indication of the fundamental
conjecture that
entropy must increase in any many-to-one mapping which is nowhere
formally proven.

Our numerical examples indicate that the inversion of $n$-body
spectra of large systems to determine the underlying two-body
potential
is subject to increasing loss of
information. However this does not mean that important information
cannot be obtained from large systems, in addition to that
which may be extracted from smaller systems.

The method remains at present restricted to cases where the
hypercentral approximation is valid, that is, for weakly
correlated systems of particles. An extension of spectral
inversion for
$n$-body spectra interacting via  forces which induce strong
correlations between particles still remains
a formidable problem for the future.

\vspace{2cm}

\noindent
This project was completed with support from the Foundation
for Fundamental Research on Matter (FOM) and the Dutch
Organisation for Scientific Research (NWO). One of us (E.J.O.G.)
wishes to thank the Technische Universit\"{a}t Wien for
hospitality and financial support for a visit.

\newpage

\section*{Figure Captions}
\begin{enumerate}
\item Figure 1: The change produced in the four-body hypercentral
and corresponding two-body potential if (a) the bound state
normalisation
constant is altered, and (b) the energy of the ground state is
altered
by $\pm 5 \%$ (indicated by dashed and dotted lines respectively)
and $\pm 10 \% $ (the dash-dotted and dash-double dotted lines).
\item Figure 2: The changes to the hypercentral potential where (a)
the bound state normalisation constant  of the ground state
is decreased by 5 \% and (b) the
ground state energy level is lowered by 5 \% for three, four,
five and
six bodies denoted by the dashed, dotted, dash-dotted and dash-double
dotted lines respectively.
\item Figure 3: The changes to the two-body potentials (a) and (b),
corresponding to
those in Figure 2 (a) and (b), as well as the total new two-body
potential (c) and (d). The solid line indicates the (change in)
potential for the two-body case.
\end{enumerate}

\newpage
\vspace{23cm}
\begin{center}
{\bf Figure 1(a)}
\end{center}
\newpage
\begin{center}
\vspace{23cm}
{\bf Figure 1(b)}
\end{center}
\newpage
\vspace{11cm}
\begin{center}
Figure 2(a)
\end{center}
\vspace{11cm}
\begin{center}
Figure 2(b)
\end{center}


\newpage

\frenchspacing
\begin{thebibliography}{99}
\bibitem{inv} Z.S. Agranovitsch and V.A. Marchenko,
{\em ``The Inverse
Problem of Scattering Theory''}, Gordon, New York (1963); \\
I.M. Gel'fand and B.M. Levitan, Izv. Akad. Nank. SSR,
ser. Mat. 109 (1951).
\bibitem{HHEM} M. Fabre de la Ripelle and J. Navarro, Ann. Phys.
{\bf 123}, 185 (1979).
\bibitem{thr} B.N.Zakhariev, Few-Body Systems {\bf 4}, 25 (1988).
\bibitem{Leeb} H.Leeb, H.Fiedeldey, E.J.O. Gavin, S.A. Sofianos and
R. Lipperheide, Few-Body Systems {\bf 12}, 55 (1992).
\bibitem{Gavin} E.J.O. Gavin, H. Fiedeldey, H. Leeb and
S.A. Sofianos, to appear in Int. J. Mod. Phys. A.
\bibitem{Richard} J.M. Richard and P. Taxil, Ann. Phys. (N.Y.)
{\bf 150},
267 (1983); Phys. Lett. {\bf 128B}, 453 (1983).
\bibitem{Fermion} M. Fabre de la Ripelle, H. Fiedeldey and
M. Wiechers,
Ann. Phys. {\bf 138}, 275 (1982); \\
R.M. Adam, S.A. Sofianos, H. Fiedeldey and M. Fabre de la Ripelle,
J. Phys. G {\bf 18}, 1365 (1992).
\bibitem{Abel} R. Courant and D. Hilbert, {\em ``Methods of
Mathematical
Physics, Vol. I''}, Interscience, New York (1962).
\bibitem{Erdelyi} A. Erd\'{e}lyi (ed.), {\em ``Tables of integral
transforms, Vol. II''}, McGraw-Hill Book Company, Inc.,
New York (1954).
\bibitem{Sukumar} C.V. Sukumar, J. Phys. A Math. Gen. {\bf 18},
2937 (1985).
\bibitem{Abraham} P.B. Abraham and H.E. Moses,
Phys. Rev. A {\bf 22},
1333 (1980).






\end{thebibliography}
\end{document}